# Investigating Quantum Spin-Textures Using Universal MJ Hamiltonians

Manish Kumar Mohanta* and Puru Jena[†]

Department of Physics, Virginia Commonwealth University, Richmond, VA 23284, USA

E-mail: *manishkmr484@gmail.com, mohantamk@vcu.edu, [†]pjena@vcu.edu

**Abstract:** This work introduces a pair of novel universal MJ spintronic models that precisely mirror the complex spin textures observed in spintronic materials. The spin-orbit coupled (SOC) Hamiltonians $\mathcal{H}_{MJ1}$ and $\mathcal{H}_{MJ2}$ reveal a range of novel and intriguing spin phenomena by modulating spin-orbit interactions. The Hamiltonian $\mathcal{H}_{MJ1}$ reshapes the existing paradigm, providing a more robust and versatile framework than the Hamiltonian $\mathcal{H}_{RD}$, with the potential to catalyze new advancements in the study of quantum materials. $\mathcal{H}_{MJ1}$ encapsulates two distinct spin textures: a unidirectional, momentum-independent persistent spin texture (PST), and a bidirectional (partial) PST. In contrast Hamiltonian $\mathcal{H}_{MJ2}$ portrays a spiral spin texture, drawing a conceptual link to the cosmological process of expansion and contraction, mirrored within a two-dimensional quantum framework. We also explore the fundamental aspects of earlier analytical models that underpin the construction of the present MJ spintronic model. The physical interpretations of these models are illustrated graphically, and the emerging spin phenomena resulting from complex SOC are elucidated using a simple vector model.

# 1. Introduction

Spintronics is an emerging field in nanotechnology that exploits the intrinsic spin degree of freedom of electrons. Spintronic devices can revolutionize data storage, computing, and quantum information by offering speed, power efficiency, and data density advantages. [1,2] Analytical spintronic models such as those of Rashba and Dresselhaus are crucial for understanding and manipulating spin-orbit coupling (SOC) effects in various materials. [3–5] These models describe the influence of spin-orbit interactions on the electronic band structure and the behavior of electron spins in semiconductors and other materials. This study makes use of the fundamentals of the Rashba and Dresselhaus spin-orbit coupling models and subsequently establishes them as the cornerstone for the MJ spintronic model.

Spin-orbit coupling is a quantum mechanical phenomenon in which an electron spins interact with its orbital motion. This interaction is pivotal in spintronics because it allows for the manipulation of spin states through electric fields, enabling spintronic devices to operate without relying on magnetic fields. The SOC is characterized by the coupling strength, which varies depending on the material and the type of spin-orbit interaction present.

The Rashba effect describes spin splitting in the electronic band structure due to SOC in systems with structural inversion asymmetry. This effect typically arises in two-dimensional electron gas systems where the inversion symmetry is broken by an external electric field or the intrinsic structure of the material. The Dresselhaus effect or bulk inversion asymmetry describes spin splitting due to SOC in crystals lacking bulk inversion symmetry, such as zinc-blende semiconductors.

In many materials, both Rashba and Dresselhaus SOC can coexist. [6,7] The interplay between these effects can lead to complex spin textures and novel phenomena such as anisotropic spin relaxation and persistent spin helix states where spin orientation remains stable over long distances. The Rashba and Dresselhaus models provide fundamental insights into the behavior of electron spins in low-dimensional systems and materials with broken inversion symmetry. These fundamental models are essential for developing new models describing novel spin phenomena, and this work is a first step.

## 2. From Conventional Analytical Models to a Universal Model

### A. 2D Rashba Model:

The 2D Rashba Hamiltonian is given by [4,8]:

$$\mathcal{H}_R = \frac{\hbar^2}{2m}(k_x^2 + k_y^2) + \alpha(\sigma_x k_y - \sigma_y k_x) \dots\dots\dots (1)$$

where $\vec{k}$ is the momentum of the electron, α corresponds to the strength of Rashba spin-orbit coupling, $m$ is the effective electron mass and $\sigma_i's$ are Pauli spin matrices. The energy eigenvalues, obtained by diagonalizing the Hamiltonian, are given by

$$E_{R_1/R_2} = \frac{\hbar^2(k_x^2+k_y^2)}{2m} \mp \alpha\sqrt{k_x^2 + k_y^2} \dots\dots\dots\dots (2)$$

The spin polarization of each eigenstate is obtained by $\vec{S} = \langle \psi_{\vec{k}} | \vec{\sigma} | \psi_{\vec{k}} \rangle$, where $\psi_{\vec{k}}$ is the eigenstate of the given Hamiltonian, as plotted in Figure 1(a). Note that the spin textures are momentum-dependent and related to the time-reversal symmetry ($\mathfrak{I}$), as observed and indicated in the figure. The overall skeletons of the spin texture are plotted sidewise for reference.

### B. 2D Dresselhaus Model: 
The 2D Dresselhaus Hamiltonian [3] is given by:

$$\mathcal{H}_D = \frac{\hbar^2}{2m}(k_x^2 + k_y^2) + \beta(\sigma_x k_x - \sigma_y k_y) \dots\dots\dots (3)$$

where $\beta$ corresponds to the strength of Dresselhaus spin-orbit coupling. The energy eigenvalues are given by

$$E_{D_1/D_2} = \frac{\hbar^2(k_x^2+k_y^2)}{2m} \mp \beta\sqrt{k_x^2 + k_y^2} \dots\dots\dots\dots (4)$$

The spin textures of both energy bands are separately plotted in Figure 1(b).

Now, we will tweak these original Hamiltonians to observe how their mathematical distinction influences the spin texture.

### C. Modified 2D Rashba Model: 
In this model, the form of the Hamiltonian presented in equation (1) is modified as follows:

$$\mathcal{H}_{MR} = \frac{\hbar^2}{2m}(k_x^2 + k_y^2) + \mathcal{M}(\sigma_x k_y + \sigma_y k_x) \dots\dots\dots (5)$$

where $\mathcal{M}$ is an SOC constant. The energy eigenvalues are given in equation (6), and the spin textures obtained from $\mathcal{H}_{MR}$ are plotted in Figure 1(c).

$$E_{MR_1/MR_2} = \frac{\hbar^2(k_x^2+k_y^2)}{2m} \mp \mathcal{M}\sqrt{k_x^2 + k_y^2} \quad \ldots\ldots\ldots\ldots (6)$$

**D. Modified 2D Dresselhaus Model (Weyl Model):** Similarly, in this model, the form of the Hamiltonian presented in equation (3) is modified as follows [9–11]:

$$\mathcal{H}_{MD} = \frac{\hbar^2}{2m}(k_x^2 + k_y^2) + \mathcal{J}(\sigma_x k_x + \sigma_y k_y) \quad \ldots\ldots\ldots (7)$$

where $\mathcal{J}$ is an SOC constant. The energy eigenvalues are given in equation (8).

$$E_{MD_1/MD_2} = \frac{\hbar^2(k_x^2+k_y^2)}{2m} \mp \mathcal{J}\sqrt{k_x^2 + k_y^2} \quad \ldots\ldots\ldots\ldots (8)$$

The spin textures obtained from $\mathcal{H}_{MD}$ are plotted in Figure 1(d).

It is important to note that while the energy dispersion relations of the Hamiltonians discussed above exhibit similar characteristics, with concentric circles forming the Fermi contours (see Supplemental Material (SM)) [12,13], their spin textures are markedly different. A key observation is that the spin textures derived from the Dresselhaus ($\mathcal{H}_D$) and modified Rashba ($\mathcal{H}_{MR}$) models are related by π/4 rotation, as illustrated in Figure 1. This relationship suggests the existence of different Hamiltonians capable of producing similar spin textures. These four models lay the groundwork for the MJ model discussed in this study.

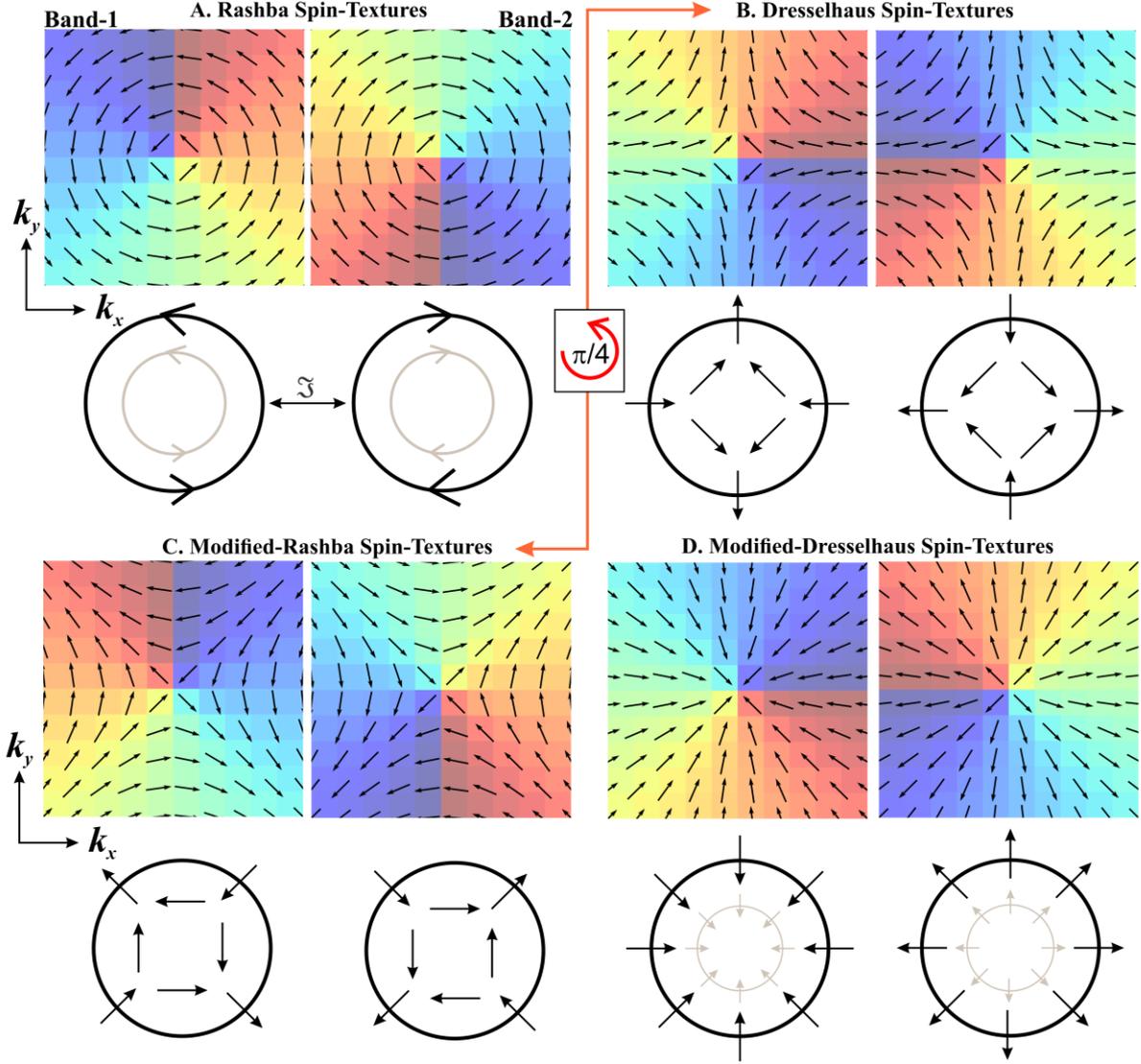

Figure 1. Spin textures obtained from the (a) Rashba Hamiltonian, (b) Dresselhaus Hamiltonian, (c) modified Rashba Hamiltonian, and (d) modified Dresselhaus Hamiltonian. Different parameters used for plotting: $\hbar = m = \alpha = \beta = \mathcal{M} = \mathcal{J} = 1$, $k_x = k_y = [-1, 1]$. The skeleton of the overall spin textures of the individual bands is plotted sidewise.

**E. 2D Quantum well model - Equal Rashba and Dresselhaus SOC Strength**

A Hamiltonian combining the linear terms of both the Rashba and Dresselhaus Hamiltonians is given by [14,15]

$$\mathcal{H}_{RD} = \frac{\hbar^2}{2m}(k_x^2 + k_y^2) + \alpha(\sigma_x k_y - \sigma_y k_x) + \beta(\sigma_x k_x - \sigma_y k_y) \quad \ldots\ldots\ldots\ldots (9)$$

The Fermi surface and spin textures derived from simplified $\mathcal{H}_{RD}$ under the conditions of equal Rashba and Dresselhaus SOC strengths $\alpha = +\beta$ are provided in Figure 2(a), and those for $\alpha =$ $-\beta$ are provided in Figure 2(b). Note that under the stringent conditions of $\alpha = \pm\beta$, a unidirectional momentum-independent spin texture, also known as persistent spin texture (PST), is observed.

As observed, the form of $\mathcal{H}_{RD}$ is complex and requires certain conditions to simplify the Hamiltonian, as discussed above. However, the simplest Hamiltonian that can exhibit PST for a 2D electron–gas system can be given by

For in-plane PST:

$$\mathcal{H}_1 = \frac{\hbar^2}{2m}\left(k_x^2 + k_y^2\right) + \sigma_x k_x \ \ldots\ldots\ldots\ldots (10)$$

$$\mathcal{H}_2 = \frac{\hbar^2}{2m}\left(k_x^2 + k_y^2\right) + \sigma_x k_y \ \ldots\ldots\ldots\ldots (11)$$

Similarly,

$$\mathcal{H}_3 = \frac{\hbar^2}{2m}\left(k_x^2 + k_y^2\right) + \sigma_y k_x \ \ldots\ldots\ldots\ldots (12)$$

$$\mathcal{H}_4 = \frac{\hbar^2}{2m}\left(k_x^2 + k_y^2\right) + \sigma_y k_y \ \ldots\ldots\ldots\ldots (13)$$

For the out-of-plane PST:

$$\mathcal{H}_5 = \frac{\hbar^2}{2m}\left(k_x^2 + k_y^2\right) + \sigma_z k_x \ \ldots\ldots\ldots\ldots (14)$$

$$\mathcal{H}_6 = \frac{\hbar^2}{2m}\left(k_x^2 + k_y^2\right) + \sigma_z k_y \ \ldots\ldots\ldots\ldots (15)$$

More details about the Hamiltonians showing out-of-plane PSTs are discussed in earlier works. [16–19] The energy eigenvalues, eigenstates, and spin polarizations of Hamiltonians showing in-plane PSTs are provided in Table 1, and their Fermi contours and spin textures are plotted in Figure 2.

Table 1 The energy eigenvalues, corresponding eigenstates, and spin polarization of Hamiltonian

| Cases | Energy eigenvalues | Eigenstates | Spin polarization |
|---|---|---|---|
| $\mathcal{H}_1 = \frac{\hbar^2}{2m}(k_x^2 + k_y^2) + \sigma_x k_x$ | $E_{11} = \frac{\hbar^2(k_x^2 + k_y^2)}{2m} - k_x$ | $\begin{pmatrix} -1 \\ 1 \end{pmatrix}$ | $\begin{pmatrix} -1 \\ 0 \\ 0 \end{pmatrix}$ |
| | $E_{12} = \frac{\hbar^2(k_x^2 + k_y^2)}{2m} + k_x$ | $\begin{pmatrix} 1 \\ 1 \end{pmatrix}$ | $\begin{pmatrix} 1 \\ 0 \\ 0 \end{pmatrix}$ |
| $\mathcal{H}_2 = \frac{\hbar^2}{2m}(k_x^2 + k_y^2) + \sigma_x k_y$ | $E_{21} = \frac{\hbar^2(k_x^2 + k_y^2)}{2m} - k_y$ | $\begin{pmatrix} -1 \\ 1 \end{pmatrix}$ | $\begin{pmatrix} -1 \\ 0 \\ 0 \end{pmatrix}$ |
| | $E_{22} = \frac{\hbar^2(k_x^2 + k_y^2)}{2m} + k_y$ | $\begin{pmatrix} 1 \\ 1 \end{pmatrix}$ | $\begin{pmatrix} 1 \\ 0 \\ 0 \end{pmatrix}$ |
| $\mathcal{H}_3 = \frac{\hbar^2}{2m}(k_x^2 + k_y^2) + \sigma_y k_x$ | $E_{31} = \frac{\hbar^2(k_x^2 + k_y^2)}{2m} - k_x$ | $\begin{pmatrix} i \\ 1 \end{pmatrix}$ | $\begin{pmatrix} 0 \\ 1 \\ 0 \end{pmatrix}$ |
| | $E_{32} = \frac{\hbar^2(k_x^2 + k_y^2)}{2m} + k_x$ | $\begin{pmatrix} -i \\ 1 \end{pmatrix}$ | $\begin{pmatrix} 0 \\ -1 \\ 0 \end{pmatrix}$ |
| $\mathcal{H}_4 = \frac{\hbar^2}{2m}(k_x^2 + k_y^2) + \sigma_y k_y$ | $E_{41} = \frac{\hbar^2(k_x^2 + k_y^2)}{2m} - k_y$ | $\begin{pmatrix} i \\ 1 \end{pmatrix}$ | $\begin{pmatrix} 0 \\ 1 \\ 0 \end{pmatrix}$ |
| | $E_{42} = \frac{\hbar^2(k_x^2 + k_y^2)}{2m} + k_y$ | $\begin{pmatrix} -i \\ 1 \end{pmatrix}$ | $\begin{pmatrix} 0 \\ -1 \\ 0 \end{pmatrix}$ |

The Fermi contour and spin states in Figure 2 indicate Hamiltonians $\mathcal{H}_{RD}$ ($\alpha = \pm\beta$), $\mathcal{H}_1$, $\mathcal{H}_2$, $\mathcal{H}_3$, $\mathcal{H}_4$, $\mathcal{H}_5$ and $\mathcal{H}_6$ to be equivalent and represent 10 possible directions of the PST, as shown in Figure 2(d). The Fermi contours displayed in Figure 2 are notably distinct from those of $\mathcal{H}_{R/D/MR/MD}$ as shown in Figure S1. The cross-section of the bands shown in Figure 2(c) reveals a spin-degenerate line node (SDLN) where the opposite spins converge along this line, resulting in a net spin of zero. [20] While the Hamiltonians $\mathcal{H}_1$, $\mathcal{H}_2$, $\mathcal{H}_3$ and $\mathcal{H}_4$ provide a basic representation of PSTs, they offer limited insights into the complexities of the 2D electron gas system.

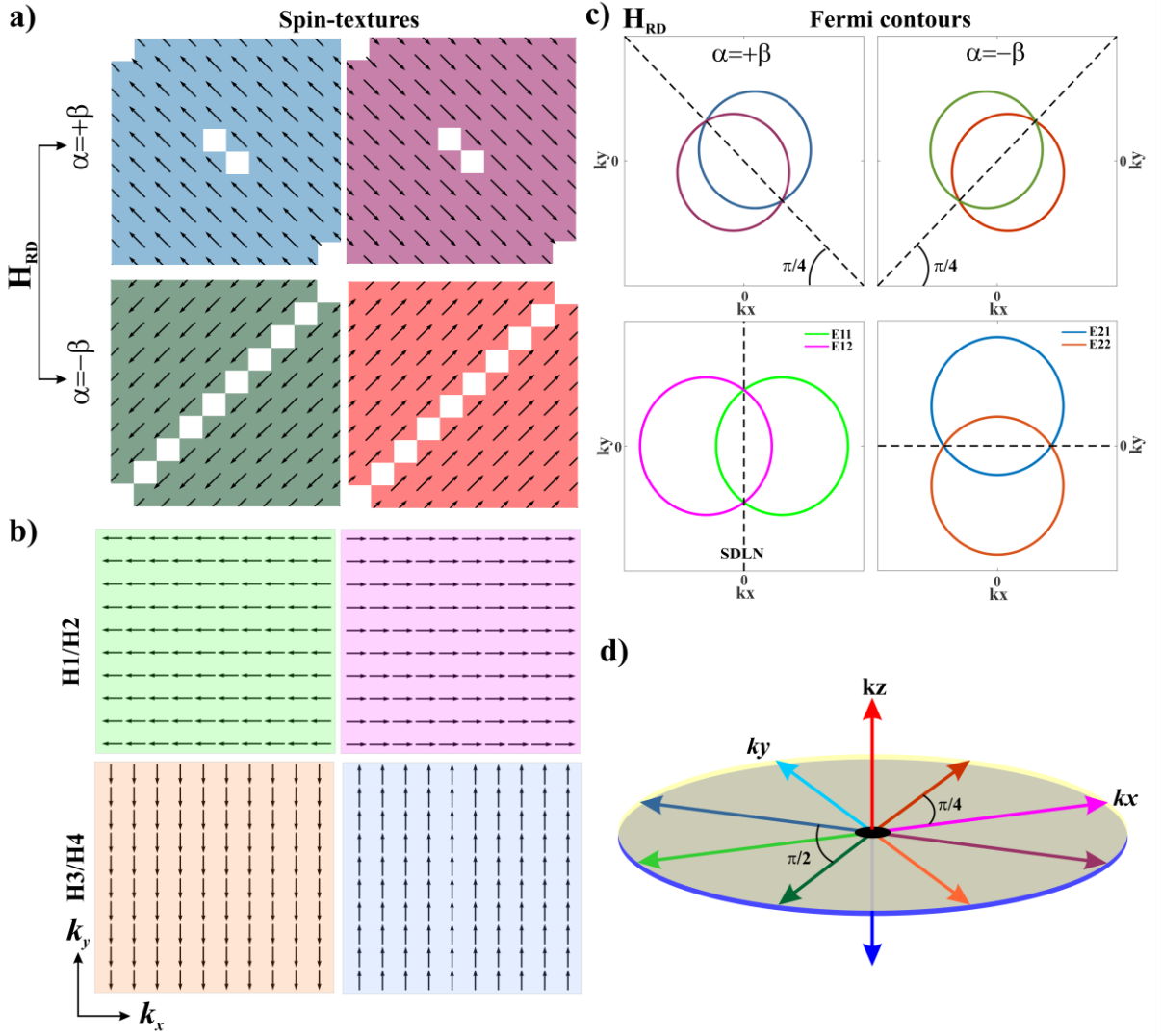

Figure 2 (a) Spin textures obtained from $\mathcal{H}_{RD}$ under the condition of $\alpha = \pm\beta$ making an angle $\pm\pi/4$ w.r.t $x/y$ axis, (b) spin textures obtained from $\mathcal{H}_{1/2/3/4}$ and parallel to $x/y$ axis, (c) Fermi contours obtained from energy eigenvalues; the dotted line represents the spin degenerate line node, and (d) 10 possible directions of the PST.

Now, to understand the physical meaning of Hamiltonian $\mathcal{H}_{RD}$, we consider two special cases: $\mathcal{H}_{RD1}$; case-1: $\alpha = \beta$; and case-2: $\beta \to \alpha$.

$$\mathcal{H}_{RD1} = \frac{\hbar^2}{2m}(k_x^2 + k_y^2) + \alpha(\sigma_x k_y - \sigma_y k_x) - \beta(\sigma_x k_x - \sigma_y k_y) \ldots\ldots\ldots (16)$$

Figure 2(a) already illustrates the spin textures for the first case; yet, the transition of spin textures as $\beta \to \alpha$ is particularly noteworthy. For clarity, Figure 3 presents the evolution of the spin texture for a single band, with the SOC parameter $\alpha$ held constant (=1) while $\beta$ varies from zero to 1. A comparison between the spin textures at $\alpha = 1; \beta = 0$ and $\alpha = 1; \beta =$ 0.5 reveals that the Dresselhaus term introduces a compressive effect on the spin polarization states. As $\beta = 0.99$, we observe the emergence of a bidirectional spin texture, referred to as partial PST. This observation is of considerable importance from a material perspective and will be discussed later.

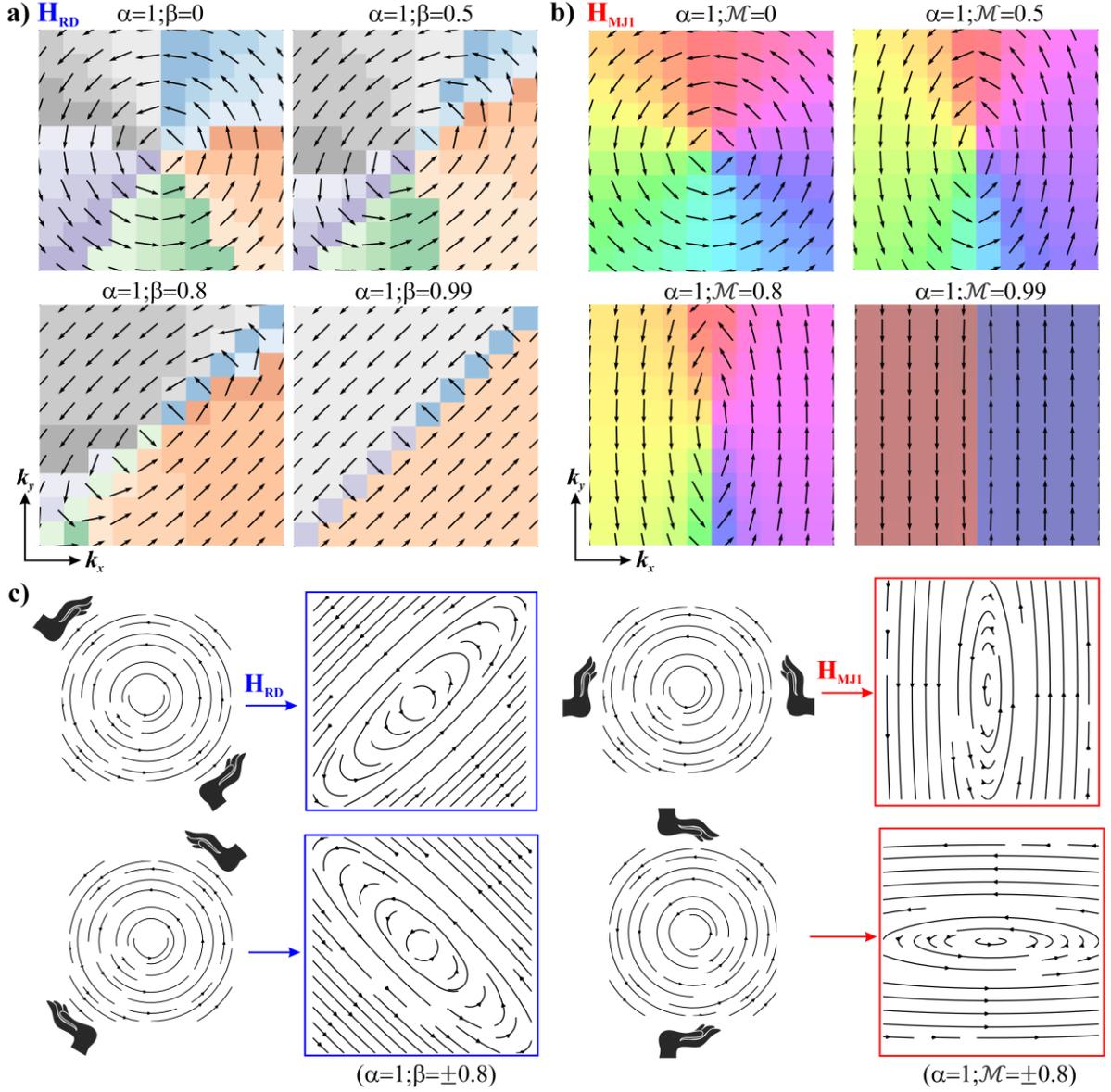

Figure 3 (a) Spin texture evolution of $\mathcal{H}_{RD}$ under $\beta \to \alpha$, (b) spin texture evolution of $\mathcal{H}_{MJ1}$ under $\mathcal{M} \to \alpha$, and (d) physical interpretation of $\mathcal{H}_{RD}$ and $\mathcal{H}_{MJ1}$ under $\beta/\mathcal{M} \to \alpha$. The different constant values represent the relative strength.

## F. Universal Hamiltonian $\mathcal{H}_{MJ1}$:

In this section, we introduce a new Hamiltonian $\mathcal{H}_{MJ1}$ given by

$$\mathcal{H}_{MJ1} = \frac{\hbar^2}{2m}\left(k_x^2 + k_y^2\right) + \alpha\left(\sigma_x k_y - \sigma_y k_x\right) + \mathcal{M}\left(\sigma_x k_y + \sigma_y k_x\right) \ldots\ldots\ldots(17)$$

where $\mathcal{M}$ is a SOC constant.

The energy eigenvalues are given by

$$E_{\pm MJ1} = \frac{\hbar^2(k_x^2+k_y^2)}{2m} \pm \sqrt{k_x^2\alpha^2 + k_y^2\alpha^2 - 2k_x^2\alpha\mathcal{M} + 2k_y^2\alpha\mathcal{M} + k_x^2\mathcal{M}^2 + k_y^2\mathcal{M}^2} \ldots\ldots\ldots (18)$$

Now, solving the Hamiltonian $\mathcal{H}_{MJ1}$ for $\alpha = -\mathcal{M}$, we obtain

$$\mathcal{H}_7 = \frac{\hbar^2}{2m}\left(k_x^2 + k_y^2\right) - 2\alpha\sigma_y k_x \ldots\ldots\ldots\ldots (19)$$

For $\alpha = +\mathcal{M}$, the Hamiltonian $\mathcal{H}_{MJ1}$ simplifies to

$$\mathcal{H}_8 = \frac{\hbar^2}{2m}\left(k_x^2 + k_y^2\right) + 2\alpha\sigma_x k_y \ldots\ldots\ldots (20)$$

The Hamiltonians $\mathcal{H}_7$ and $\mathcal{H}_8$ take a form similar to that given in Table 1. For $\mathcal{H}_7$ and $\mathcal{H}_8$, the spin polarization states are along the $\pm y$ and $\pm x$ axes, respectively.

The Hamiltonian $\mathcal{H}_{MJ1}$ incorporates both the Rashba term and modified Rashba SOC term, with the latter introducing a $\pi/4$ rotation to the spin texture relative to $\mathcal{H}_{RD}$. This Hamiltonian is pivotal in spintronics, as it addresses a long-standing problem observed in spintronic materials.

The importance and outcomes of the Hamiltonian $\mathcal{H}_{MJ1}$ are as follows:

(i) The physical interpretation reveals that Hamiltonians $\mathcal{H}_{RD}$ and $\mathcal{H}_{MJ1}$ are fundamentally equivalent, despite their differing forms.

(ii) The Hamiltonian $\mathcal{H}_{MJ1}$ proves to be more tractable mathematically when $\alpha = \pm\mathcal{M}$ compared to $\mathcal{H}_{RD}$ under $\alpha = \pm\beta$. Additionally, the Hamiltonians $\mathcal{H}_1, \mathcal{H}_2, \mathcal{H}_3$ and $\mathcal{H}_4$ can be derived directly from $\mathcal{H}_{MJ1}$. Under the condition $\alpha = \pm\mathcal{M}$, the spin polarization states are along the $\pm x$ and $\pm y$ directions.

(iii) A fascinating result of this Hamiltonian emerges when $\mathcal{M} \to \alpha$. Specifically when $\alpha = 1; \mathcal{M} = 0.99$, the spin textures depicted in Figure 3(b) precisely replicate the partial PST of bulk BiInO$_3$, as reported by Tao and Tsymbal. [21] This suggests that the conditions $\alpha = \pm\beta$ and $\alpha = \pm\mathcal{M}$ are indeed critical for manifesting PST, and even a minimal deviation, such as 99% relative strength fails to exhibit such remarkable properties. Therefore, selecting the appropriate material is of paramount importance for experimental realization. A quick comparison can be made with the spin textures of previous reports [22,23], CsBiNb$_2$O$_7$ [24], and CsPbBr$_3$ [25].

(iv) The variation in the overall spin texture pattern is as follows: circle ($\alpha = 1$; $\beta/\mathcal{M} = 0$) → ellipse → bidirectional PST → unidirectional PST ($\alpha = 1$; $\beta/\mathcal{M} = 1$).

(v) The Hamiltonian $\mathcal{H}_{MJ1}$ is elegantly structured, relying on just two parameters $\alpha$, which remains constant and $\mathcal{M}$ which is varied. This simplicity makes it superior to the multivariable Hamiltonian derived from the *k.p* model by Tao and Tsymbal.

(vi) The Hamiltonian $\mathcal{H}_{MJ1}$ Hamiltonian introduces a universal framework for a 2D electron gas system, effectively distinguishing between PSTs and partial PSTs under varying conditions, while accurately replicating the spin textures observed in spintronic materials.

## G. Hamiltonian $\mathcal{H}_{MJ2}$ and Spiral Spin Texture

The Hamiltonian combining the Rashba and modified Dresselhaus terms is given by

$$\mathcal{H}_{MJ2} = \frac{\hbar^2}{2m}\left(k_x^2 + k_y^2\right) + \alpha\left(\sigma_x k_y - \sigma_y k_x\right) + \mathcal{J}\left(\sigma_x k_x + \sigma_y k_y\right) \quad \ldots\ldots (21)$$

Diagonalizing the Hamiltonian $\mathcal{H}_{MJ2}$, the energy eigenvalues are obtained and are given by;

$$E_{MJ21/MJ22} = \frac{\hbar^2(k_x^2+k_y^2)}{2m} \mp \sqrt{k_x^2\alpha^2 + k_y^2\alpha^2 + k_x^2\mathcal{J}^2 + k_y^2\mathcal{J}^2} \quad \ldots\ldots (22)$$

The eigenstates of $\mathcal{H}_{MJ2}$ are $\begin{pmatrix} -\frac{i\sqrt{(k_x^2+k_y^2)(\alpha^2+\mathcal{J}^2)}}{(k_x+ik_y)(\alpha+i\mathcal{J})} \\ 1 \end{pmatrix}$ and $\begin{pmatrix} \frac{i\sqrt{(k_x^2+k_y^2)(\alpha^2+\mathcal{J}^2)}}{(k_x+ik_y)(\alpha+i\mathcal{J})} \\ 1 \end{pmatrix}$.

The spiral spin texture obtained from $\mathcal{H}_{MJ2}$ is given in Figure 4(a). Such spin textures have been observed in earlier reports. [26,27]

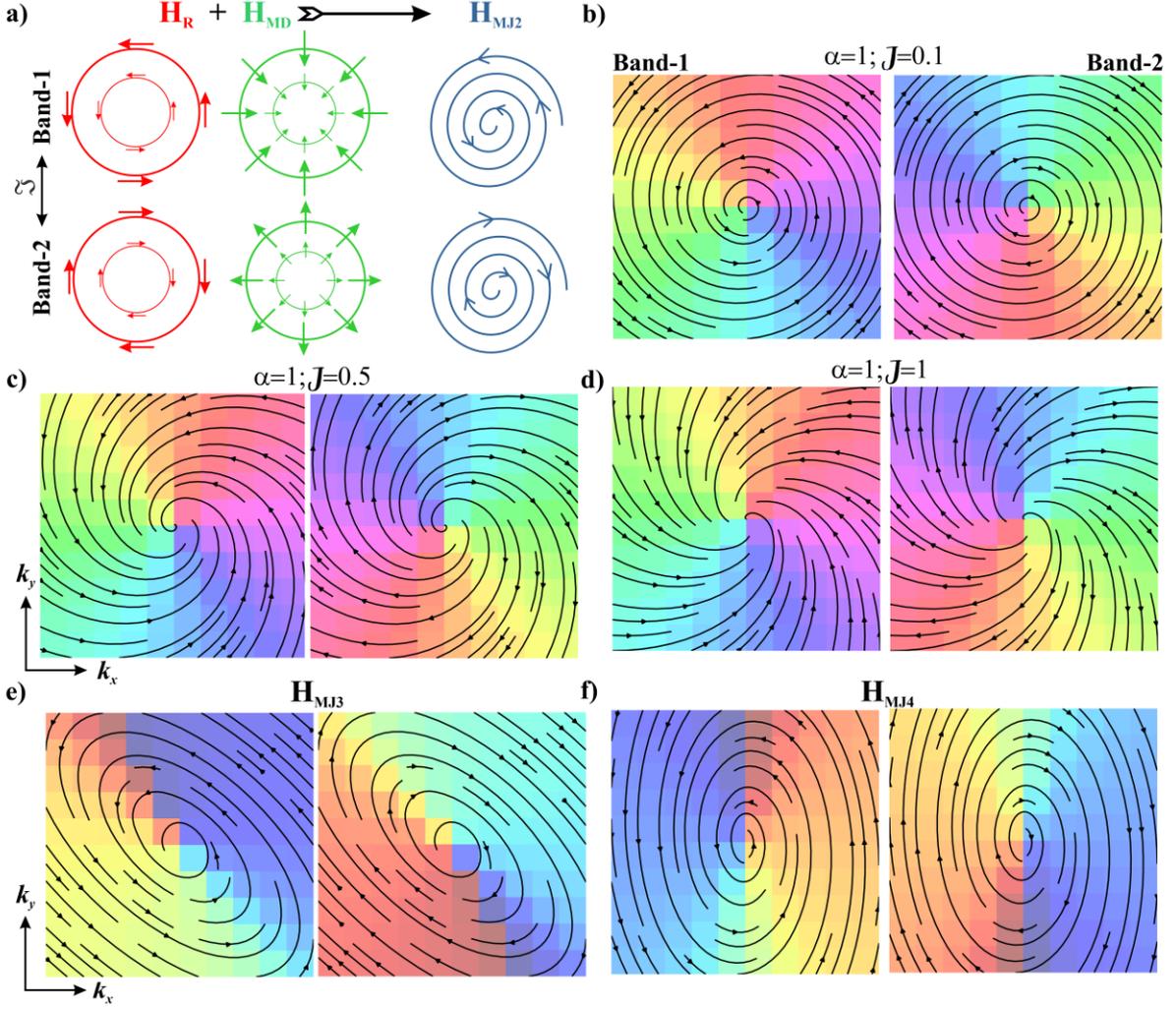

Figure 4 (a) The overall representation of spin textures for $\mathcal{H}_{MJ2}$, (b-d) evolution of spin textures of $\mathcal{H}_{MJ2}$ under different relative SOC strengths, and (e-f) spin textures obtained from anonymous Hamiltonians $\mathcal{H}_{MJ3}$ and $\mathcal{H}_{MJ4}$.

## H. Observations and correlation

### H1. Spinors as Simple Vectors

**Observation 1:** Referring to the Hamiltonian $\mathcal{H}_{RD}$, it is noteworthy in Figure 2(a) that under the condition of $\alpha = +\beta$, spin states are given by $\begin{pmatrix} 1 \\ \mp e^{-i\pi/4} \end{pmatrix}$ and for $\alpha = -\beta$, the lower spin component acquires an additional factor of $(-i)$. [15] The spin orientation forms an angle $\pi/4$ w.r.t $\pm x$ axis. But why does it adopt this specific angle? To address this, one should consider that the angle $\pi/4$ is a significant angle in vector algebra, representing the direction of the resultant vector when two equal vectors, perpendicular to each other, are added or subtracted. This observation prompts the question: can spin dynamics be linked to a simple vector model?

**Observation 2:** From Figure 1, one can observe/assume that the spin-orbit coupled Hamiltonian forms a vector field that influences the spin dynamics/states. Now, when two SOC terms are included in a Hamiltonian, as in $\mathcal{H}_{RD}$, $\mathcal{H}_R$ and $\mathcal{H}_D$ generate two different vector fields. How do these two fields interact when they are combined?

The best example for illustration is under the $\alpha = \pm\beta$ condition. We focus on the spin texture of only one band of the following Hamiltonians:

$$\mathcal{H}_{RD1} = \mathcal{H}_R - \mathcal{H}_D = \frac{\hbar^2}{2m}(k_x^2 + k_y^2) + \alpha(\sigma_x k_y - \sigma_y k_x) - \beta(\sigma_x k_x - \sigma_y k_y)\ldots\ldots (23)$$

and

$$\mathcal{H}_{RD2} = \mathcal{H}_R + \mathcal{H}_D = \frac{\hbar^2}{2m}(k_x^2 + k_y^2) + \alpha(\sigma_x k_y - \sigma_y k_x) + \beta(\sigma_x k_x - \sigma_y k_y)\ldots\ldots (24)$$

The skeletons of spin textures $\mathcal{H}_R$ and $\mathcal{H}_D$ are shown in Figure 1. In the following example, we simply perform vector addition and subtraction, assuming that spin states are vectors in a Cartesian coordinate system. The red/green colors are used to represent the Rashba/Dresselhaus vectors, and the blue color represents the resultant vector. The resultant vectors are $\pi/4$ clockwise w.r.t. the positive and negative x-axes, as shown in Figure 5 (II). These are the exact angles observed in Figure 2(a). However, Figure 5(II) has mixed resultant vectors and requires certain symmetry constraints to obtain Figure 5(III).

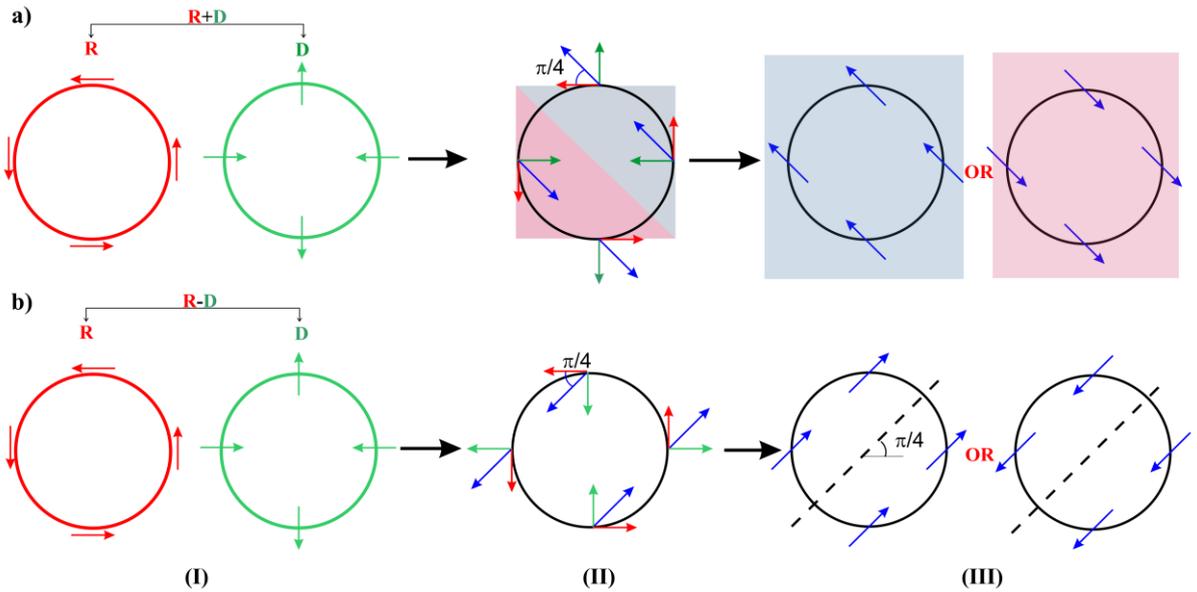

Figure 5 Simple vector addition and subtraction of two vectors R and D represented by red and green colors, respectively, whose magnitudes are equal to each other. (a-b) Vector algebra analogous to $\mathcal{H}_R \pm \mathcal{H}_D$.

**Observation 3:** Note that the vector field is zero along the dotted line in Figure 5(b-III). Does this phenomenon also manifest in the spin texture plot shown in Figure 2(a) under the condition $\alpha = -\beta$? Additionally, could this vector model be extended to the spin textures derived from the Hamiltonian $\mathcal{H}_{MJ2}$ in Figure 4(a)? A more complex spin texture, as shown in Figure 4(e-f), can also be generated using the MJ model, leaving the formulation of the Hamiltonian $\mathcal{H}_{MJ3}$ and $\mathcal{H}_{MJ4}$ as an exercise for the readers.

## H2. Deciphering the Quantum Universe

**Observation 4:** The Hamiltonian $\mathcal{H}_{MJ2} = \frac{\hbar^2}{2m}(k_x^2 + k_y^2) + \alpha(\sigma_x k_y - \sigma_y k_x) + \mathcal{J}(\sigma_x k_x + \sigma_y k_y)$ offer more physical interpretation rather than spiral spin textures for quantum materials. In a different context, the spiral texture is an optimal representation of the expansion/contraction of the quantum universe in a 2D plane. The transition from a circular to a spiral pattern requires an external force/factor pointing inward/outward, as depicted in Figure 4(a). As illustrated in Figure 4, it provides a bird's-eye view of the expansion/contraction of the quantum universe, which can be analogous to the expansion of our universe. A comparison across Figures 4(b-d) shows that the rate of expansion/contraction is strongly influenced by the constant $\mathcal{J}$.

By setting the second term in $\mathcal{H}_{MJ2}$ to zero, i.e., for $\mathcal{J} = 0$, the spiral spin pattern simplifies to a circular form. This observation is intriguing, especially when contrasted with the spiral and expanding nature of the universe. Could the second term be responsible for the spiral structure and expansion? Moreover, what was the aerial view pattern of the universe before expansion began? Are these phenomena connected? Does this Hamiltonian capture the essential 2D pattern if the universe were to contract instead of expand? Analogous concepts can be extended to elliptical patterns (refer to Figure 4(e-f)).

Despite offering an approximate and multivariable Hamiltonian for observed spin textures in line with group theory and crystal symmetry, the *k.p* theory necessitates further simplification. [28] In contrast, this study adopts a straightforward and simplified approach, offering a comprehensive analysis that underscores the importance and physical interpretation of various SOC Hamiltonians.

**Conclusion:**

In this work, two universal spin-orbit coupled Hamiltonians $\mathcal{H}_{MJ1}$ and $\mathcal{H}_{MJ2}$ are proposed. The Hamiltonian $\mathcal{H}_{MJ1}$ generalized the difference between the full PST and partial PST and exactly reproduced the partial PST observed in spintronic materials, whereas $\mathcal{H}_{MJ2}$ characterized spiral spin textures. Although Hamiltonians $\mathcal{H}_{MJ1}$ and $\mathcal{H}_{MJ2}$ are simple and systematic, they effectively capture the complexity of spin textures emerging from sophisticated spin-orbit interactions. Furthermore, the spiral spin textures obtained from the Hamiltonian $\mathcal{H}_{MJ2}$ can be intriguingly linked to the expansion/contraction of the universe. The models in this study capture the spin phenomena that emerge when multiple SOC terms are incorporated into a Hamiltonian framed by a straightforward vector model. This study also revealed that alternative forms of Hamiltonian can produce equivalent spin textures, such as $\mathcal{H}_D \leftrightarrow \mathcal{H}_{MR}$, $\mathcal{H}_{RD} \leftrightarrow \mathcal{H}_{MJ1}$(unconditionally) and $\mathcal{H}_{RD}\ (\alpha = \pm\beta) \leftrightarrow \mathcal{H}_{1/2/3/4/5/6} \leftrightarrow \mathcal{H}_{MJ1}(\alpha = \pm\mathcal{M})$. Moreover, this work provides a broad analysis and opens a new avenue for creating Hamiltonians that capture intricate spin textures, presenting a fresh way of interpreting spin phenomena within complex Hamiltonians.

**Acknowledgments**

MKM and PJ acknowledge financial support from the U.S. Department of Energy, Office of Basic Energy Sciences, Division of Materials Sciences and Engineering under Award No. DE-FG02-96ER45579. Resources of the National Energy Research Scientific Computing (NERSC) Center supported by the Office of Science of the U.S. Department of Energy under Contract No. DE-AC02-05CH11231 are also acknowledged. The authors extend their acknowledgment to the High-Performance Research Computing (HPRC) core facility at Virginia Commonwealth University for providing supercomputing resources.